\documentclass[%
 article,%aip,
 amsmath,amssymb,
 reprint,%
 nofootinbib,
 draft
]{revtex4-1}

\usepackage{graphicx}
\usepackage{dcolumn}% Align table columns on decimal point
\usepackage{bm}% bold math
\usepackage{color}

\begin{document}

\preprint{}

\title[An Exact Solution of a Generalization of the Rabi Model]{An Exact Solution of a Generalization of the Rabi Model}

\author{F. Moolekamp}
\email{fmooleka@pas.rochester.edu.}
\affiliation{Unversity of Rochester, Rochester, NY 14627,USA}

\date{\today}

\begin{abstract}
There has been renewed theoretical interest recently in the Rabi model due to
Braak's analytical solution and introduction of a new criterion for
integrability. We focus not on the integrability of the system but
rather why it is solvable in the first place. We show that the Rabi
model is the limiting case of a more general finite dimensional system
by use of a contraction and suggest that this is the reason for it's
solvability, which still applies in the case of non-integrable but
solvable variations.
\end{abstract}

\maketitle %\maketitle must follow title, authors, abstract and \pacs

\section{Introduction}

In classical mechanics the terms integrable and solvable are often 
used interchangably, implying that a system that is integrable is also said 
to be solvable while a non-integrable system exibits chaos. 
In this sense it became clear in the early twentieth century that most classical
dynamical systems are not analytically solvable, for example the perturbation
series for the three body problem is only convergent
in certain regions of the phase space. Later work by Kolmogorov, Arnold
and Moser \cite{KAM} established that this region covers a large volume
for small perturbations but has a complicated fractal structure. Thus,
chaos and instability are still possible for small bodies in nearly
Keplerian orbits and the solar system appears stable because such
bodies were either kicked out or fell into the Sun or Jupiter. 

An analogous understanding still does not exist in the case of quantum mechanics.
The superficial observation that the problem to be solved is linear
(the Schrodinger equation) misses the point that the Hilbert spaces
of most systems of interest are infinite dimensional: a linear problem
in infinite dimensions has many of the analytical subtleties of finite
dimensional non-linear ordinary differential equations \cite{vonNeumann}. One way to
approach the question of solvability of a quantum system is to ask
if it has sensible finite dimensional truncations which are exactly
solvable.

Recently Braak \cite{Braak} analytically solved the Rabi model \cite{Rabi},
with Hamiltonian
\begin{equation}
H_{R}=\omega a^{\dagger}a+\frac{\Delta}{2}\sigma_{3}+g\sigma_{1}\left(a+a^{\dagger}\right),\label{eq:H_R}
\end{equation}
which can be thought of as a two level atom with energy levels $\pm\Delta/2$
coupled to a quantized field with frequency $\omega$. As Braak mentions, the general opinion
was that this system is not solvable because the Hamiltonian is its
only conserved quantity. A second conserved quantity would allow the
Hamiltonian to be brought to diagonal form. For example, \cite{TwoLevel}
in the Jayne-Cummings model $Q=a^{\dagger}a+\frac{1}{2}\sigma_{3}$
is conserved. In the Rabi model, although $Q$ is not conserved, it
only changes by a bounded amount: $|\Delta Q| \leq 1$. 
We can see this by defining $\mid q\rangle$ as a basis in which $Q$ 
is block diagonal, so that $\langle q'|H|q \rangle=0$ 
if $\Delta Q=|q-q'| \, > 1$. This makes the
Hamiltonian block-tridiagonal and we solve it in terms of continued
fractions. Unlike Braak, who was interested in the integrability
of the system, we ask another fundamental question ``What makes
a quantum mechanical system solvable?''

To view this problem from a different perspective we look at the Hamiltonian
\begin{equation}
H_{L}=\omega L_{3}+\Delta R_{3}+gL_{1}R_{1},\label{eq:H_L}
\end{equation}
where $L$ and $R$ are angular momentum matrices with magnitude $l$
and $r$ respectively and we have absorbed a factor of 2 into $g$. This looks eerily similar to the Rabi model
and in section \ref{sec:Contraction} we use a contraction of the algebra to show that in the limit
as $l\rightarrow\infty$ with $r=\frac{1}{2}$, (\ref{eq:H_L}) becomes
the Rabi Hamiltonian (\ref{eq:H_R}). The advantage of studying this new, more general
Hamiltonian, is that unlike the Rabi model it has finite dimension.

To find the matrix elements of $H_{L}$ we use the basis states
\begin{equation}
\mid\psi_{L}\rangle=\mid l,m_{l},\, r,m_{r}\rangle=\mid l,m_{l}\rangle\otimes\mid r,m_{r}\rangle \label{eq:states}
\end{equation}
where $m_{l}=-l,-l+1,...,l$ and $m_{r}=-r,-r+1,...,r$ are the azimuthal
components of angular momenta in the $L_3$, $R_3$ basis. This allows us to write the Hamiltonian
as a block tridiagonal matrix 
\begin{equation}
H_{L}=\left(\begin{array}{ccccc}
A_{-l} & B_{-l+1} & 0 & \cdots & 0\\
B_{-l+1} & A_{-l+1} & B_{-l+2} & \cdots & 0\\
0 & B_{-l+2} & A_{-l+2} & \ddots & 0\\
0 & 0 & \ddots & \ddots & B_{l}\\
0 & 0 & \cdots & B_{l} & A_{l}\end{array}\right)\label{eq:H_L_matrix}
\end{equation}
with 
\begin{align}
A_{k} & =  k\omega I_{r}+\Delta R_{3}\nonumber \\
B_{k} & =  \sqrt{l(l+1)-k(k-1)}gR_{1},\label{eq:A,B}
\end{align}
where $I_{r}$ is an $r$ dimensional identity matrix.

\section{Continued fractions and tridiagonal matrices\label{sec:Continued-fractions}}

The general theory of analytical continued fractions was developed
by Stieltjes in the late $19^{th}$ century while studying divergent
power series\cite{Wall}. These continued fractions are of the form
\begin{equation}
f(z)=\frac{1}{z+a_{1}-\frac{b_{1}^{2}}{z+a_{2}-\frac{b_{2}^{2}}{z+a_{3}-\dots}}}\label{eq:cf}
\end{equation}
and are intimately connected with ordinary tridiagonal matrices of
the form 
\begin{equation}
A=\left(\begin{array}{ccccc}
a_{0} & b_{1} & 0 & 0 & .\\
c_{1} & a_{1} & b_{2} & 0 & .\\
0 & c_{2} & a_{2} & b_{3} & .\\
0 & 0 & c_{3} & a_{3} & .\\
. & . & . & . & .
\end{array}\right).\label{eq:tridiag}\end{equation}
This connection can be seen by defining
\begin{align}
\Delta_{-1}(z) & \equiv  1\nonumber \\
\Delta_{0}(z) & =  a_{0}-z\nonumber \\
\Delta_{k}(z) & =  \left(a_{k}-z\right)\Delta_{k-1}(z)-b_{k}c_{k}\Delta_{k-2}(z)\label{eq:Delta_k}
\end{align}
which gives us a continued fraction 
\begin{align}
S_{k}(z) & =  \frac{\Delta_{k}(z)}{\Delta_{k-1}(z)}\nonumber \\
 & =  \left(a_{k}-z\right)-\frac{b_{k}c_{k}}{S_{k-1}(z)}\label{eq:S_k_tri}
\end{align}
whose roots are the eigenvalues of the matrix. If however, $A$ is
an infinite dimensional matrix, the roots of $S_{k}$ represent the $k^{th}$
approximation to the eigenvalues of $A$. When $a_{k}$ are real and
$b_{k}c_{k}>0$ the function $S_{k}$ is a Sturm sequence, meaning
the zeros are real and the roots of $S_{k-1}$ are the poles of $S_{k}$.
This gives us an easy way to calculate higher approximations to the
eigenvalues: between any two pairs of poles of $S_{k}$ is an eigenvalue.

This tells us that as long as $S_{k}$ is a convergent continued fraction,
even if the matrix it represents is infinite dimensional, it can still
be solved to the desired level of precision. More importantly since
the convergence of such continued fractions has been well established 
for over a hundred years this method gives a true check as to whether or not
diagonalizing increasingly larger matrices will converge to the eigenvalues
of an infinite dimensional matrix\cite{Wall}.

\section{Eigenvalues of block tridiagonal matrices\label{sec:Eigenvalues-of-block}}

The usefulness of tridiagonal form to prove convergence and the block
tridiagonal form of $H_{L}$ prompts us to ask {}``is it also possible
to develop relations similar to (\ref{eq:Delta_k}) and (\ref{eq:S_k_tri})
for block tridiagonal matrices?'' The answer is, in some cases, yes.
Using the transfer matrix method of Molinari \cite{Molinari} we can
find the eigenvalues of an $(n+1)\times(n+1)$ block tridiagonal matrix
\begin{equation}
M=\left(\begin{array}{ccccc}
A_{0} & B_{1}\\
B_{1} & A_{1} & B_{2}\\
 & B_{2} & A_{2} & \ddots\\
 &  & \ddots & \ddots & B_{n}\\
 &  &  & B_{n} & A_{n}
\end{array}\right),\label{eq:M}\end{equation}
where $A_{k},B_{k}$ are $m\times m$ square matrices, and $A_{k}$
is of the form $A_{k}^{'}-zI_{m}$, where $z$ are the eigenvalues.
In this case $\textrm{det}M=0$, allowing us to set 
\begin{equation}
M\Psi=\left(\begin{array}{ccccc}
A_{0} & B_{1} & 0 & 0 & .\\
C_{1} & A_{1} & B_{2} & 0 & .\\
0 & C_{2} & A_{2} & B_{3} & .\\
0 & 0 & C_{3} & A_{3} & .\\
. & . & . & . & .
\end{array}\right)
\left(\begin{array}{c}
\psi_{0}\\
\psi_{1}\\
\psi_{2}\\
\psi_{3}\\
.\end{array}\right)=0\label{eq:M*Psi}\end{equation}
where $\Psi$ is a null vector with components $\psi_{k}\in\mathbb{C}^{m}$,
giving us a set of equations 
\begin{align}
A_{0}\psi_{0}+B_{1}\psi_{1} & =  0\nonumber \\
B_{k+1}\psi_{k+1}+A_{k}\psi_{k}+C_{k}\psi_{k-1} & =  0\nonumber \\
A_{n}\psi_{n}+C_{n}\psi_{n+1} & =  0.\label{eq:A,B,C:psi}
\end{align}
This can be written recursively as 
\begin{equation}
\left[\begin{array}{c}
\psi_{k+1}\\
\psi_{k}
\end{array}\right]=
\left[\begin{array}{cc}
-B_{k+1}^{-1}A_{k} & -B_{k+1}^{-1}C_{k}\\
I_{m} & 0
\end{array}\right]
\left[\begin{array}{c}
\psi_{k}\\
\psi_{k-1}\end{array}\right],\label{eq:psi_k}\end{equation}
which defines the transfer matrix 
\begin{eqnarray}
T_{k} & = & \left[\begin{array}{cc}
-B_{k+1}^{-1}A_{k} & -B_{k+1}^{-1}C_{k}\\
I_{m} & 0\end{array}\right]T_{k-1}\nonumber \\
T_{n} & = & \left[\begin{array}{cc}
A_{n} & C_{n}\\
I_{m} & 0\end{array}\right]\left[\begin{array}{cc}
-B_{n}^{-1}A_{n-1} & -B_{n}^{-1}C_{n-1}\\
I_{m} & 0\end{array}\right]\dots\label{eq:T_n} \nonumber \\
 &  & \times\left[\begin{array}{cc}
-B_{1}^{-1}A_{0} & -B_{1}^{-1}\\
I_{m} & 0\end{array}\right].\end{eqnarray}
If we define the top left element of $T_{k}$ as \begin{equation}
T_{k,11}=-B_{k+1}^{-1}A_{k}T_{k-1,11}-B_{k+1}^{-1}C_{k}T_{k-2,11},\label{eq:Tk}\end{equation}
since $\psi_{n+2}=0$ and $\psi_{-1}=0$ \begin{equation}
\left[\begin{array}{c}
0\\
\psi_{n+1}\end{array}\right]=T_{n}\left[\begin{array}{c}
\psi_{1}\\
0\end{array}\right]\label{eq:T_n*psi}\end{equation}
so $\textrm{det}T_{n,11}=0$ is the same as $\textrm{det}M=0$, defining
the eigenvalue equation.

\section{Spectrum of $H_{L}$\label{sec:Spectrum-of HL}}

For the most general form of $H_{L}$, integer values of $r$ correspond
to singular matrices that have no inverse, so the above method can
be modified to the form of Salkuyeh\cite{Salkuyeh}. It is possible
to solve $H_{L}$ for any half integer spin but the most elegant case,
which is also interesting because of it's ties to the Rabi model $H_{R}$,
is for $r=\frac{1}{2}$. In this case 
\begin{eqnarray}
R_{i} & = & \frac{1}{2}\sigma_{i}\nonumber \\
A_{k} & = & kI+\frac{\Delta}{2}\sigma_{3}\nonumber \\
B_{k} & = & C_{k}=g\sqrt{l(l+1)-k(k-1)}\sigma_{1}\nonumber \\
  & = & b_{k} \sigma_{1}\nonumber \\
B_{k}^{-1} & = & b_{k}^{-1}\sigma_{1} \label{eq:R,A,B,BI}\end{eqnarray}
where we have absorbed a factor of $\frac{1}{2}$ into $g$ and defined $b_{k}=g\sqrt{l(l+1)-k(k-1)}$. This
gives us the eigenvalue equation $\textrm{det }\left(H_{L}-zI\right)=0=\textrm{det}\left(T_{11}(z)\right),$
where the transfer matrix can be simplified by multiplying each $2\times2$
matrix in $T_{k}$ by $b_{k}$
to give 
\begin{align}
T_{k} & =  \left[\begin{array}{cc}
\sigma_{1}\left(A_{k}-zI\right) & b_{k} I\\
-b_{k} & 0\end{array}
\right]T_{k-1}\nonumber \\
T_{k,11} & =  \sigma_{1}A_{k} T_{k-1,11}-b_{m}^2 T_{k-2,11}\nonumber \\
 & =  \left(\left(k\omega-z\right)\sigma_{1}-i\frac{\Delta}{2}\sigma_{2}\right)T_{k-1,11}-B_{k}^{2}T_{k-2,11}. \label{eq:T_k,11}
\end{align}
Due to the special property of Pauli matrices $\sigma_{i}\sigma_{j}=\delta_{ij}+i\epsilon_{ijk}\sigma_{k}$
we see that if we define \begin{equation}
S_{m}=T_{m,11}T_{m-1,11}^{-1}\label{eq:S_m def}\end{equation}
we have the recursive matrix \begin{equation}
S_{m}=\left(m\omega-z\right)\sigma_{1}-i\Delta\sigma_{2}-B_{k}^{2}S_{m-1}^{-1},\label{eq:S_m}\end{equation}
which is of the form \begin{equation}
S_{m}=\left(\begin{array}{cc}
0 & a(z)\\
b(z) & 0\end{array}\right),\label{eq:Sm, a,b}\end{equation}
with inverse \begin{equation}
S_{m}^{-1}=\left(\begin{array}{cc}
0 & \frac{1}{b(z)}\\
\frac{1}{a(z)} & 0\end{array}\right).\label{eq:SmI}\end{equation}
This allows us to write $S_{m}$ in matrix form \begin{equation}
S_{m}=\left(
\begin{array}{cc}
0 & a_{m}^{+}-\frac{b_{m}^{2}}{a_{m-1}^{-}-\frac{b_{m-1}^{2}}{a_{m-2}^{+}-\ldots}}\\
a_{m}^{-}-\frac{b_{m}^{2}}{a_{m-1}^{+}-\frac{b_{m-1}^{2}}{a_{m-2}^{-}-\ldots}} & 0\end{array}\right),\label{eq:S_m matrix}
\end{equation}
where $a_{m}^{\pm}=m\omega-z\pm\frac{\Delta}{2},\,$
and by setting $S_{m}=0$ we obtain two terminating continued fractions
\begin{eqnarray}
S_{m,\pm}(z) & = & m\omega-z\mp(-1)^{m}\frac{\Delta}{2}-\frac{b_{m}^{2}}{S_{m-1,\pm}(z)},\label{eq:S_m_cf}
\end{eqnarray}
where $m=-l,-l+1,...,l$, whose zeros are the eigenvalues of the $H_{L}$.
Because these are finite dimensional matrices, calculating the roots
of $S_{l,+}$ gives the even parity spectrum while the roots of
$S_{l,-}$ give the odd parity spectrum.

\section{Contraction of $H_{L}$ to $H_{R}$\label{sec:Contraction}}

We could of course perform the same procedure on the Rabi model $H_{R}$
to find it's eigenvalues in a similar way but a deeper connection
between the two systems can be seen by performing a singular change
of basis. It was discovered by Inonu and Wigner \cite{Inonu} that
a transformation of this type changes one Lie Algebra into another
using a process called contraction. We briefly summarize Gilmore's\cite{Gilmore} 
description of a different class of contractions to show the relationship 
between (\ref{eq:H_R}) and (\ref{eq:H_L}).

A Lie algebra defined by the basis vectors $X_{i}$ is closed under
commutation, so the commutators \begin{equation}
[X_{i},X_{j}]=C_{ij}^{\,\, k}X_{k}\label{eq:[X,X]}\end{equation}
are contained in the algebra. The structure constants $C_{ij}^{\,\, k}$
completely determine the algebra, however it is possible to perform
a change of basis transformation \begin{equation}
Y_{i}=M_{i}^{\, j}X_{j}\label{eq:Y_i}\end{equation}
where the new structure constant $C_{ij}^{'\, k}$ becomes \begin{equation}
C_{ij}^{'\, k}=\left(M^{-1}\right)_{i}^{\, l}\left(M^{-1}\right)_{j}^{\, m}C_{lm}^{\,\, n}M_{n}^{\, k}\label{eq:C_{ij}^k}\end{equation}
due to a non-singular transformation. If we allow the transformation
to be parameter dependent, where \begin{eqnarray}
Y_{i} & = & M_{i}^{\, j}\left(\epsilon\right)X_{j}\nonumber \\
C_{ij}^{\, k} & = & C_{ij}^{\,\, k}\left(\epsilon\right)\label{eq:epsilon}\end{eqnarray}
the structure constant often converges to a new Lie Algebra if 
$C_{ij}^{\,\, k}\left(\epsilon\right)$ becomes singular in the limit as $\epsilon \rightarrow \infty$.

One representation of angular momentum is the compact unitary group
$U(2)$, which is spanned by the operators $J_{3},J_{\pm},J_{0}$,
($J_{0}$ is the identity) which correspond to the commutation relations
\begin{eqnarray}
[J_{3},J_{\pm}] & = & \pm J_{\pm}\nonumber \\
{}[J_{+},J_{-}] & = & 2J_{3}\nonumber \\
{}[J_{0},\mathbf{J}] & = & 0.\label{eq:[Ji,Jj]}\end{eqnarray}
We can now change basis to the Heisenberg group $H_{4}$ by using
\begin{equation}
\left(\begin{array}{c}
h_{+}\\
h_{-}\\
h_{3}\\
h_{0}\end{array}\right)=\left(\begin{array}{cccc}
c\\
 & c\\
 &  & 1 & \frac{1}{2c^{2}}\\
 &  &  & 1\end{array}\right)\left(\begin{array}{c}
J_{+}\\
J_{-}\\
J_{3}\\
J_{0}\end{array}\right)\label{eq:contraction}\end{equation}
which gives us \begin{eqnarray}
\left[h_{3},h_{\pm}\right] & = & \pm h_{\pm}\nonumber \\
\left[h_{+},h_{-}\right] & = & 2c^{2}h_{3}-h_{0}\nonumber \\
\left[h_{0},\mathbf{h}\right] & = & 0\label{eq:[hi,hj]}\end{eqnarray}
and in the limit as $c\rightarrow0$ \begin{eqnarray}
\left[h_{3},h_{\pm}\right] & = & \pm h_{\pm}\nonumber \\
\left[h_{+},h_{-}\right] & = & -h_{0}=-I\label{eq:[h,h]}\end{eqnarray}
which satisfy the same commutations as the single mode photon operators
\begin{eqnarray}
\left[N=a^{\dagger}a,a\right] & = & -a\nonumber \\
\left[N=a^{\dagger}a,a^{\dagger}\right] & = & a^{\dagger}\nonumber \\
\left[a^{\dagger},a\right] & = & -1.\label{eq:[a,ad]}\end{eqnarray}
We can now identify (in the limit as $c\rightarrow0$) \begin{eqnarray}
h_{3} & = & N\nonumber \\
h_{+} & = & a^{\dagger}\nonumber \\
h_{-} & = & a.\label{eq:h=a}\end{eqnarray}
To see how the basis states change we first operate $h_{3}$ on the
angular momentum state $\mid j,m\rangle$ to get \begin{equation}
h_{3}\mid j,m\rangle=\left(J_{3}+\frac{1}{2c^{2}}J_{0}\right)\mid j,m\rangle=\left(m+\frac{1}{2c^{2}}\right)\mid j,m\rangle.\label{eq:h state}\end{equation}
The ground state of this system corresponds to $m=-j$, so the $n^{th}$
state is $n=j+m$. In order for the limit to be well defined we require
\begin{equation}
\lim_{c\rightarrow 0}\left(m+\frac{1}{2c^{2}}\right)=\lim_{c\rightarrow 0}\left(n-j+\frac{1}{2c^{2}}\right)\label{eq:lim1}
\end{equation}
to also be well defined. We have already equated $h_{3}$ with the
number operator, so the requirement becomes \begin{equation}
\lim_{c\rightarrow 0}\left(-j+\frac{1}{2c^{2}}\right)=0\label{eq:lim2}\end{equation}
telling us that $2jc^{2}=1,$ or in other words as $c\rightarrow0$,
$j\rightarrow\infty$, and \begin{equation}
\lim_{\underset{J\rightarrow\infty}{c\rightarrow0}}h_{3}\mid j,m\rangle=n\mid\infty,n\rangle.\label{eq:lim3}\end{equation}
Similarly, we see that \begin{eqnarray}
a^{\dagger}\mid n\rangle & = & \lim_{c\rightarrow0}\, cJ_{+}\mid j,m\rangle\nonumber \\
 & = & \lim_{c\rightarrow0}\,\sqrt{j(j+1)-m(m+1)}\mid j,m+1\rangle\nonumber \\
 & = & \lim_{c\rightarrow0}\,\sqrt{(1-c^{2}n)(n+1)}\mid j,m+1\rangle\nonumber \\
 & = & \sqrt{n+1}\mid n+1\rangle.\label{eq:lim state}\end{eqnarray}
So a contraction on $H_{L}$ changes (\ref{eq:S_m_cf}) to
\begin{eqnarray}
S_{k,\pm}(z) & = & k\omega-z\mp(-1)^{k}\frac{\Delta}{2}-\frac{g^{2}k}{S_{k-1,\pm}(z)},\label{eq:Hr spectrum}\end{eqnarray}
$k=0,1,2,3,\dots$ Its zeros give us the spectrum of the Rabi model $H_{R}$. As discussed
in section \ref{sec:Continued-fractions}, there is a rapidly convergent
algorithm to find the zeros, which takes advantage of the fact that
there is a zero of $S_{k,\pm}(z)$ in between two of its poles, which
are simply the zeroes of the previous approximation, $S_{k-1,\pm}(z)$.
Thus we can limit the search for each zero to these intervals, increasing
the size $k$ of the matrix in each step.

\section{Conclusion}

We show that the Rabi model can be thought of as the limit of a sequence
of finite dimensional block tridiagonal Hamiltonians, each of which
can be solved by a continued fraction method. The solvability of the
Rabi model can thus be understood as due to this finite dimensional
truncation and the existence of approximate conservation laws (selection
rules for transition matrix elements) that ensure block tridiagonality, 
explaining why even a Hamiltonian with a broken symmetry can be solved \cite{Braak}. 
Conversely, we should expect that systems which do not
allow convergent finite dimensional approximations exhibit quantum
chaos. We hope to construct such an example in a later publication.

\begin{acknowledgments}
I would like to thank S.G. Rajeev for his insights and guidance in
writing this paper as well as J. Eberly, S. Agarwal, A. Kar, C. Broadbent
and X. Qian for discussions.
\end{acknowledgments}

\bibliography{paper}

\end{document}